\documentclass[aps,physrev,twocolumn,superscriptaddress]{revtex4-2}

\usepackage{xcolor}
\usepackage{graphicx}
\usepackage{dcolumn}
\usepackage{bm}

\usepackage{lineno}
\usepackage{amsmath}
\usepackage{subcaption}
\usepackage{float}
\usepackage{lipsum}
\usepackage{ragged2e}
\usepackage{braket}

\begin{document}


\title{All-dielectric metasurface polarization scrambler for imaging applications}


\author{Edith Hartmann}
\email[]{edith.hartmann@fresnel.fr}
\author{Michel Lequime}

\affiliation{Aix Marseille Univ, CNRS, Centrale Med, Institut Fresnel, Marseille, France}

\author{Matthieu Castelnau}
\affiliation{CNES, 18 Ave Edouard Belin, F-31400 Toulouse, France}

\author{Hélène Krol}
\affiliation{CILAS, Pole Alpha Sud St Mitre, 600 Ave Roche Fourcade, F-13400 Aubagne, France.}

\author{Myriam Zerrad}
\affiliation{Aix Marseille Univ, CNRS, Centrale Med, Institut Fresnel, Marseille, France}

\begin{abstract}
The partial polarization of the Earth's radiation can be the source of radiometric errors for space instruments with polarization-sensitive components. It is therefore necessary to have a device capable of depolarizing the light before it enters the instrument, i.e. a polarization scrambler. Current scramblers used in optical payloads generally comprise combinations of prisms that enable spatial depolarization. However, polarization insensitivity comes at the cost of an inevitable degradation in imaging performance. Here, we present a device based on an all-dielectric metasurface using anisotropic scatterers capable of generating multiple polarization states by varying their orientation angle. Our new scrambling solution allows a massive reduction in the integrated degree of polarization and thus the spatial depolarization of any incident linear polarization, while reducing the impact on its image quality and allowing easier integration into the instrument design.
\end{abstract}


\maketitle

\section{Introduction}
\label{Introduction}
Earth radiance has the property of being partially polarized. This partial polarization is linear and depends on phenomena such as Rayleigh scattering from aerosol clouds or reflection from the ground. Unfortunately, this uncontrolled state of polarization can be the source of radiometric errors when some parts of an instrument used to observe or monitor the Earth from space have their own polarization signature. Gratings are an example of this. It is therefore necessary to have a device capable of depolarizing the light before it enters the instrument, i.e. a polarization scrambler \cite{bezy_polarization_2017}.

Several types of scramblers have been developed to address this question. The first, known as the Lyot depolarizer, combines two quartz wedge crystals aligned at an angle of $45^\circ$ between the two optical axes, and can achieve spectral depolarization under broadband illumination \cite{a_p_loeber_depolarization_1982}. However, the proper operation of this type of depolarizer is compromised when working with narrow bandwidths. This limitation can be overcome by using scramblers based on temporal or spatial depolarization. Since passive components are preferred when considering the requirements of space applications, most of the scramblers used in optical payloads involve combinations of prisms that allow spatial depolarization by imposing polarization-dependent phase delays that vary across the pupil \cite{bezy_polarization_2017,loesel_microcarb_2018}. The Dual-Babinet design is a prime example of this. It is used in various spectrometry missions, including the MicroCarb mission, which we will use as our case study. MicroCarb is a space instrument developed by the French space agency CNES with the aim of accurately monitoring the concentration of greenhouse gases in the Earth's atmosphere. The optical design of the instrument includes a spectrometer and a Dual-Babinet polarization scrambler to ensure polarization insensitivity \cite{pasternak_microcarb_2017,loesel_microcarb_2018}. The main limitation of these prism-based scramblers is that the scrambling process inevitably degrades image quality. The incident light passing through the prisms is separated into several polarized beams diffracted in different directions. The polarization scrambler being placed at the instrument entrance pupil, this give rise to a "diamond effect", i.e. a splitting of the instrument's point spread function into multiple spots on the detector \cite{loesel_microcarb_2018,chipman_analysis_1990}. Recall that the point spread function (PSF) is a measure of the imaging performance of an optical system, and represents the image of a single point-like object.
In addition to PSF division, spot intensities vary depending on the scene's polarization state, which can lead to geolocation errors in measurements \cite{loesel_microcarb_2018,chipman_analysis_1990}. These side effects have been managed to ensure the smooth operation of various space missions using this type of prism-based scrambler. In the case of the MicroCarb mission, the diamond effect produced by the scrambler is sufficiently weak compared to the area of one sounding spot ($25\text{ km}^2)$ \cite{loesel_microcarb_2018}. However, this could become a real limitation for the next generation of satellites, which require extensive polarization insensitivity and high-resolution performance.
As no solution is currently optimal, it is interesting to explore new methods and find a better trade-off between depolarization and image quality.

Metasurfaces are planar devices that are spatially structured at the wavelength scale. They have recently emerged as a new technology capable of pushing the boundaries of traditional optical components by allowing sub-wavelength control of light properties \cite{genevet_recent_2017}, enabling ultra-thin and high quality components such as meta-lenses \cite{khorasaninejad_metalenses_2016,chen_broadband_2019}, miniaturized spectrometers \cite{billuart_towards_2021}, or even full Stokes polarization cameras \cite{rubin_matrix_2019}.

It has already been shown that metasurfaces happen to be a promising solution for both temporal \cite{jie_electronically_2022} and spatial depolarization \cite{schau_polarization_2012,wang_ultra-compact_2020}. 
Our work aims to go beyond what has been done by presenting an all-dielectric metasurface that allows spatial depolarization of any linear incident polarization, with quasi-perfect transmission and a reduced impact on image quality. Our work is based on the use of subwavelength arrays of high refractive index dielectric nanoresonators (scatterers) with identical anisotropic shape but different orientations.
Anisotropic scatterers introduce shape birefringence and phase jumps. Their subwavelength dimensions are optimized so that the scatterers behave similarly to half-wave plates \cite{rubin_polarization_2021,balthasar_mueller_metasurface_2017,hu_all-dielectric_2020}. The scatterers will thus rotate linear polarization states like a half-wave plate would. The idea of our work is then to use a large number of these scatterers on a metasurface and to vary the orientations of their fast axis. In this way, it will be possible to generate a wide range of local output polarization states across the device, allowing any linear incident polarization to be scrambled.

The performance of our device is quantified by the reduction of the integrated Degree of Polarization (DoP) and the preservation of the PSF. The goal of this work is to develop a new polarization scrambling solution for imaging applications that overcomes the current limitations of prism-based scramblers.

\section{Unit cell design}
\label{sec:UnitCellDesign}
\begin{figure}[H]
	\begin{center}
		\includegraphics[width=\linewidth]{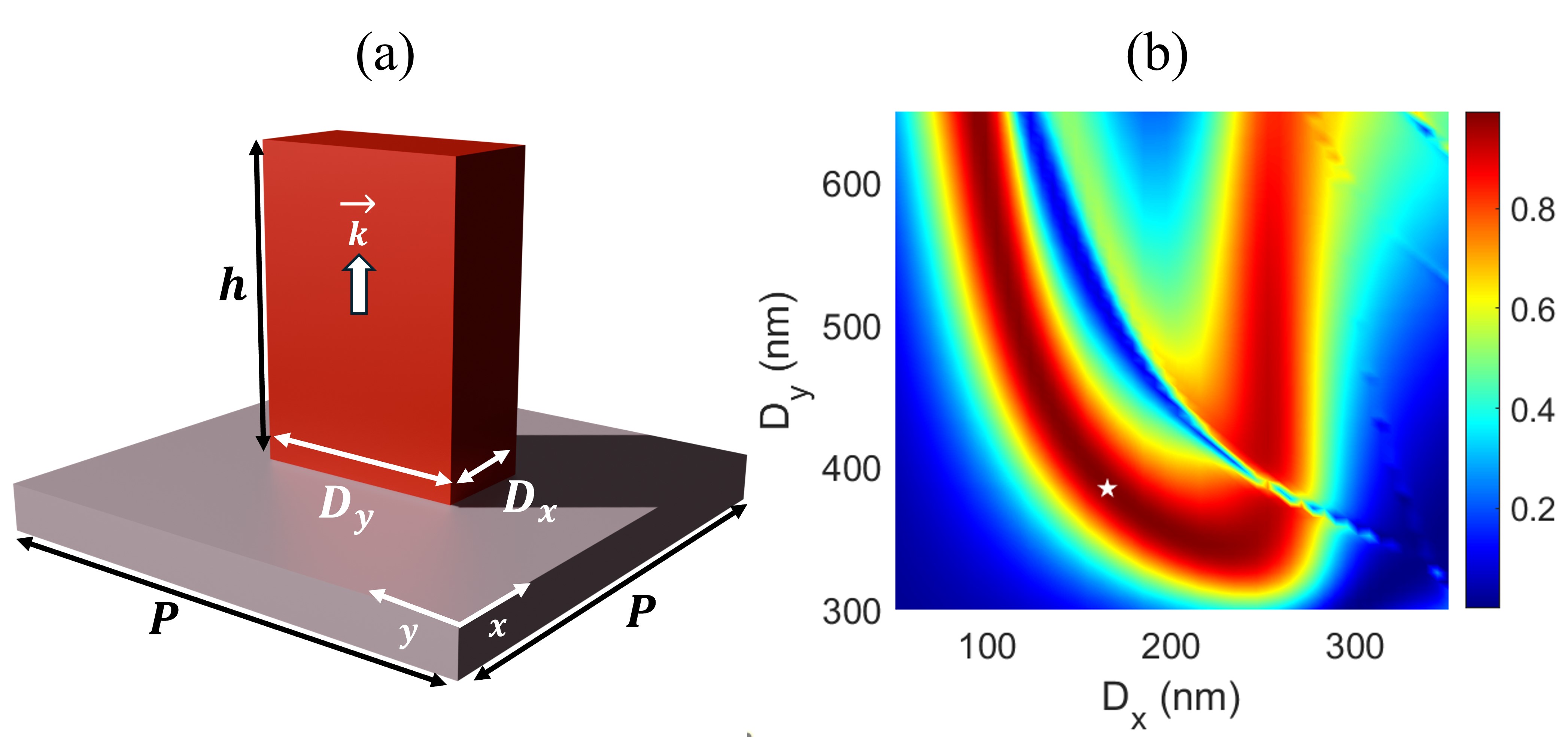}
		\caption{(a) Unit cell. - (b) Circular polarization conversion efficiency $\eta$ calculated at a wavelength of 1605 nm as a function of the width $D_x$ and the length $D_y$ of the pillar (the side $P$ of the unit cell is equal to 800 nm and the height $h$ of the pillar is equal to 1000 nm). The optimized values are represented by the white star.}
		\label{fig:Unit_cell}
	\end{center}
\end{figure}

The first step is to define the unit cell (or meta-atom) that will be used to build the metasurface. This unit cell is here formed by a rectangular pillar of silicon (length $D_y$, width $D_x$, height $h$) located in the center of an elementary square area of a silica substrate (side $P$), as shown in Fig.\ref{fig:Unit_cell}a. 
The two materials used here (silicon [Si] and silica [SiO$_2$]) are transparent at the design wavelength ($\lambda = 1605$ nm), and have a high refractive index contrast ($n_{\text{Si}}=3.74,\thinspace n_{\text{SiO$_2$}} = 1.44$). This improves the efficiency of our device, minimizes losses and allows it to operate in transmission mode. The anisotropic nature of the scatterer (rectangular shape) induces form birefringence, which is particularly well suited for polarization control applications \cite{rubin_polarization_2021,hu_all-dielectric_2020}. This unit cell can be represented by a diagonal Jones matrix \textbf{J}
\begin{equation}
\textbf{J} = 
	\begin{pmatrix}
	t_{xx} & 0\\
	0 & t_{yy} \\
	\end{pmatrix}
	\label{eq:Jones_Matrix}
\end{equation}
where $t_{xx}$ and $t_{yy}$ are the complex transmission coefficients along the two axes of the pillar. The form birefringence allows the meta-atom to behave like a waveplate, by inducing a phase difference between its two axes. The value of this phase difference can be tuned by playing with the dimensions of the pillars \cite{rubin_polarization_2021}. We optimize the dimensions of the pillars to make their behavior as close as possible to that of a half-wave plate. This is achieved by maximizing the circular polarization conversion efficiency $\eta$, defined as follows
\begin{equation}
\eta = \bigg|\textbf{J} \left|L\rangle\langle R\right|\bigg|^2
\label{eq:eta}
\end{equation}
where $\big|L\rangle$ and $\big|R\rangle$ describe the left- and right-handed circular polarization states, respectively, i.e.
\begin{equation}
\big|L\rangle=\frac{1}{\sqrt{2}}
\begin{pmatrix}
1\\
i\\
\end{pmatrix}
\enskip\text{;}\enskip
\langle R|=\frac{1}{\sqrt{2}}
\begin{pmatrix}
1 & -i\\
\end{pmatrix}
\end{equation}

The modeling of the optical properties of a pillar is carried out by means of a rigorous coupled wave analysis (RCWA) performed by a solver (Lumerical) commercialized by ANSYS. For each set of geometric parameters, the solver considers a 2D periodic replication of the unit cell and calculates the Jones matrix \textbf{J}. Then, the circular polarization conversion efficiency $\eta$ is calculated using equation (\ref{eq:eta}). Figure \ref{fig:Unit_cell}b shows an example of the results when the two optimized parameters at a wavelength of 1605 nm are the width $D_x$ and the length $D_y$ of the pillar, while keeping the height $h$ of the pillar and the side $P$ of the unit cell constant at 1000 nm and 800 nm, respectively.

The optimized values of the two parameters are $D_x=165$ nm and $D_y=385$ nm (represented by a white star on Fig.\ref{fig:Unit_cell}b), corresponding to a circular polarization conversion efficiency $\eta$ of $0.9925$ at the design wavelength $\lambda = 1605$ nm.

\section{Building the metasurface}
\label{sec:BuildingtheMetasurface}
\begin{figure}[H]
	\begin{center}
		\includegraphics[width=0.95\linewidth]{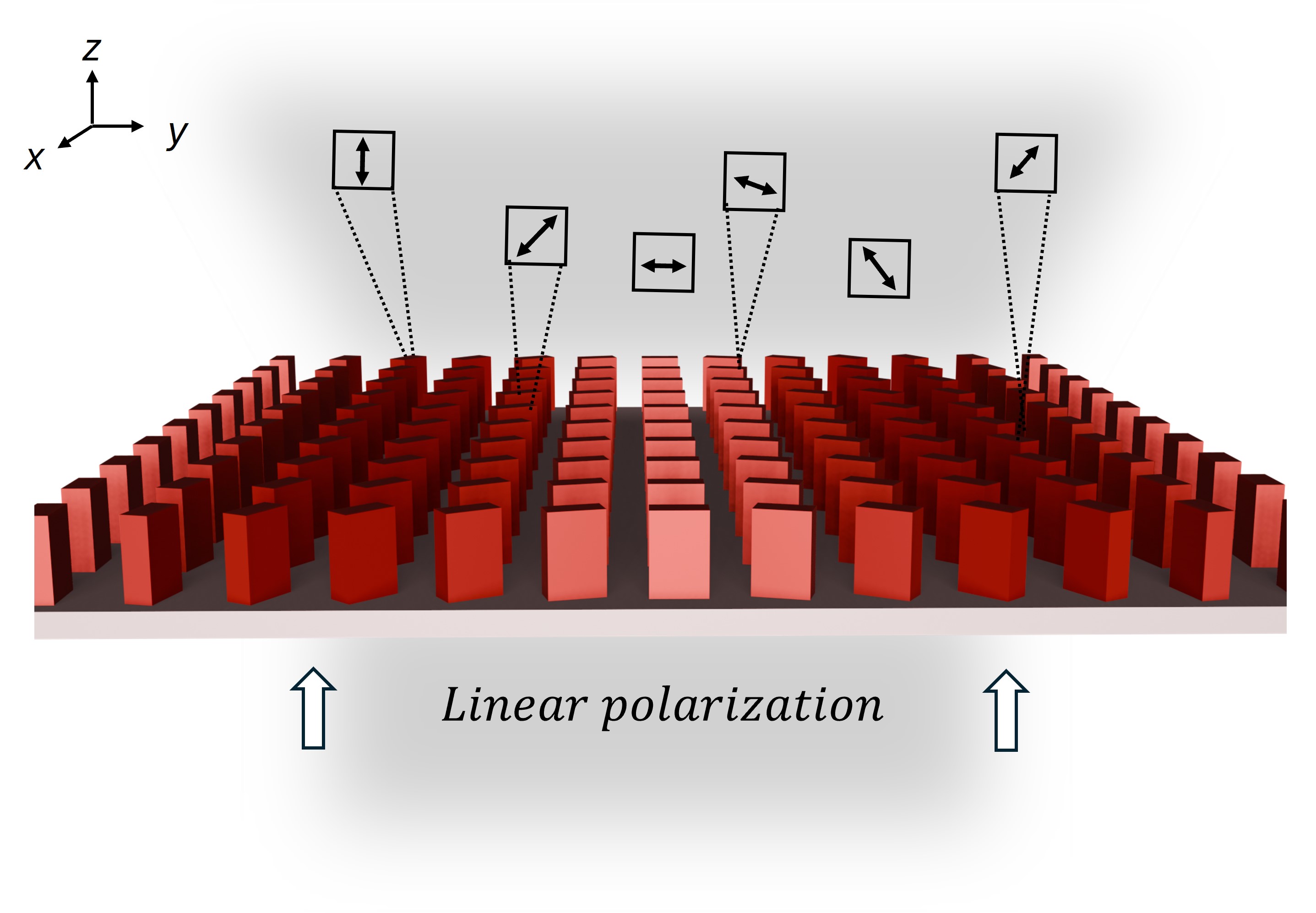}
		\caption{Operating principle: Linear incident polarization is converted into various local polarization states.}
		\label{fig:Metasurface}
	\end{center}
\end{figure}

All scatterers at the surface of the polarization scrambler have the same geometric parameters, and the only difference between them is their orientation angle $\theta(x,y)$ varying in space across the component (see Fig. \ref{fig:Metasurface}).

The global response of the metasurface is then built by combining the local contributions of each pillar.

\subsection{Local contribution of a rotated pillar}
The half-wave plate behavior of the pillars has been optimized for an orientation of $\theta=0^\circ$ (see Section \ref{sec:UnitCellDesign}). However, this behavior is slightly affected when considering other orientation angles. Therefore, to account for these slight variations, we will compute the local contribution of rotated pillars by performing RCWA simulations dedicated to each angle $\theta$. This approach is explained and justified in more detail in the appendix.

\subsection{Fourier Matrix Formalism}
\label{sec:FourierMatrixFormalism}
\begin{figure}[!htpb]
	\begin{center}
		\includegraphics[width=0.95\linewidth]{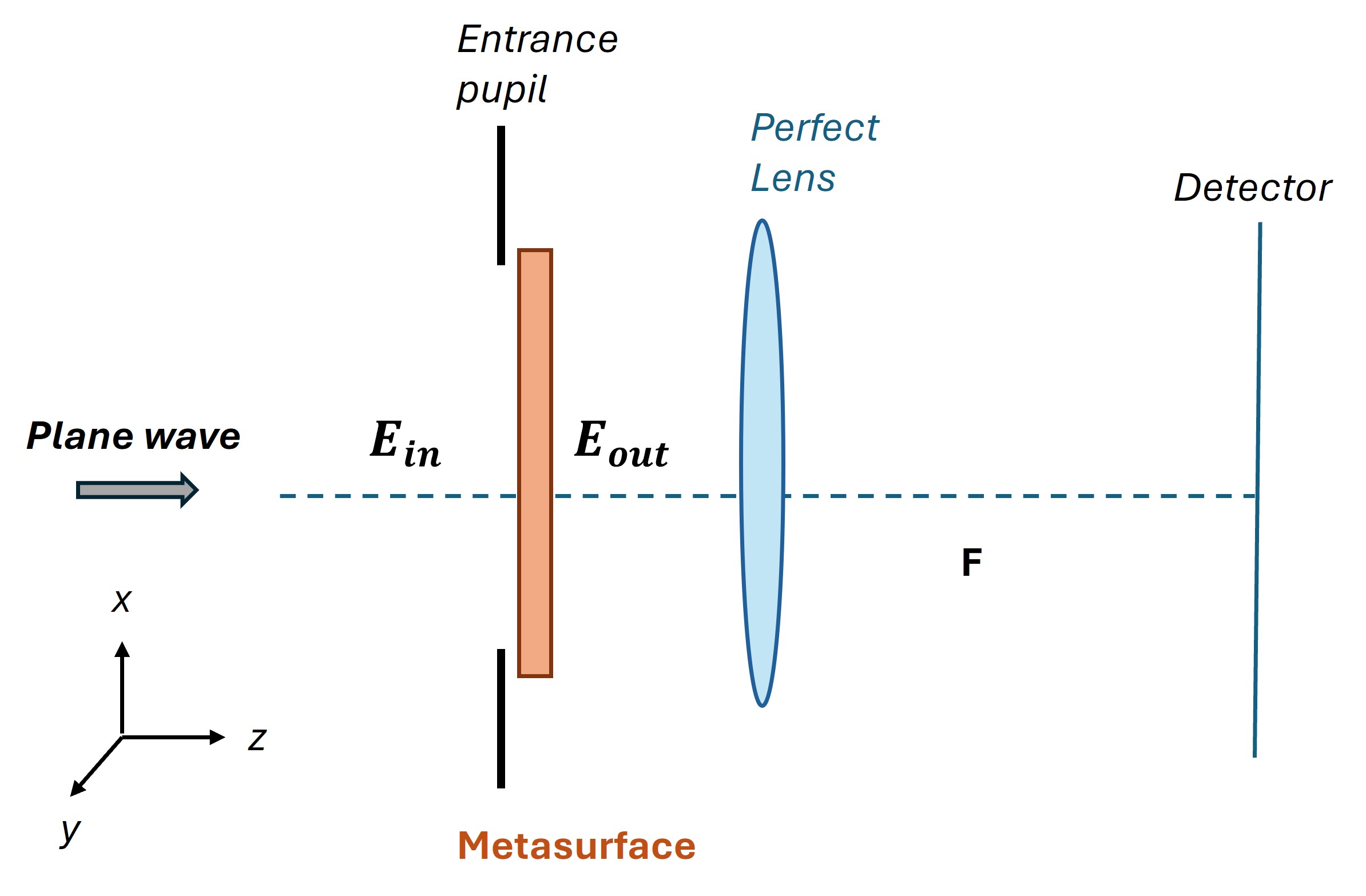}
		\caption{Optical system considered.}
		\label{fig:OpticalSystem}
	\end{center}
\end{figure}

To evaluate the impact of the metasurface on the image quality, we compute the resulting PSF of a perfect imaging system with our device at the entrance pupil (Fig.\ref{fig:OpticalSystem}).

However, a number of considerations must be taken into account when dealing with a beam that has spatially varying polarization states. As with our metasurface design, the polarization states vary from pillar to pillar, generating an output beam with spatially varying polarization states. This renders the usual scalar formalism unsuitable. These polarization variations can be accounted for using a vectorial extension of diffraction theory known as the Fourier matrix formalism \cite{rubin_matrix_2019,rubin_polarization_2021}. 

The contributions of the meta-atoms are considered local in the near field. Thus, the metasurface is represented as a 2x2 matrix, the Jones pupil \textbf{J}$(x, y)$, which varies in space and takes the local orientation of the pillars into account. This matrix is only defined in the spatial domain $\Omega$ corresponding to the entrance pupil. The field directly exiting the metasurface $\ket{E_{out}(x,y)}$ is calculated discretely after each meta-atom. We assume that the incident electric field, represented by the Jones vector $\ket{E_{inc}}$, has a uniform linear polarization state and propagates along the z-axis. Please note that the following formalism requires a paraxial approximation.
\begin{equation}
\ket{E_{out}(x,y)}=\begin{bmatrix}
    E_{out_{X pol}}(x,y)\\
    E_{out_{Y pol}}(x,y)
\end{bmatrix}=\textbf{J}(x,y) \ket{E_{inc}}
\label{eq:JonesPupil}
\end{equation}
Then, the field amplitude distribution in the detection plane at the point with coordinates $(x_d,y_d)$ is proportional to the Fourier transform of the field exiting the metasurface
\begin{equation}
\ket{E_d(f_x,f_y)}= \iint\ket{E_{out}(x,y)}\thinspace e^{-2i\pi(xf_x+yf_y)}\thinspace dxdy
\label{eq:Diffraction}
\end{equation}
for the spatial frequencies defined by 
\begin{equation}
f_x=\frac{x_d}{\lambda f}\text{ ; }f_y=\frac{y_d}{\lambda f}
\notag
\end{equation}
By combining equations (\ref{eq:JonesPupil}) and (\ref{eq:Diffraction}), we can establish a matrix relationship linking the field amplitude distribution in the detection plane to the incident polarization state
\begin{equation}
\ket{E_d(f_x,f_y)}=\textbf{ARM}\ket{E_{inc}}
\end{equation}
\textbf{ARM} is called the amplitude response matrix and is defined as follows \cite{breckinridge_polarization_2015}: 
\begin{equation}
\textbf{ARM}=
\begin{pmatrix}
\mathcal{F}[J_{XX}(x,y)] &\mathcal{F}[J_{XY}(x,y)]\\
\mathcal{F}[J_{YX}(x,y)] &\mathcal{F}[J_{YY}(x,y)]
\end{pmatrix}
\label{eq:ARM}
\end{equation}
where $\mathcal{F}$ is a 2D spatial Fourier transform. The resulting PSF can be obtained by calculating the square modulus of the field amplitude distribution in the detection plane \cite{breckinridge_polarization_2015}
\begin{equation}
\text{PSF}(x_d,y_d)\propto\left|\left|\ket{E_d\left(\frac{x_d}{\lambda f},\frac{y_d}{\lambda f}\right)}\right|\right|^2
\label{eq:PSF}
\end{equation}

Since the components of $\textbf{J}(x,y)$ are complex quantities, it is interesting to note that spatial linear variations in their phase induce a positional shift in their contribution to the PSF \cite{breckinridge_polarization_2015}. This will be discussed in more detail later in the article.

Since the metasurface is placed at the entrance pupil, the distribution of pillar orientations $\theta(x,y)$ across the component directly impacts the PSF. To illustrate this phenomenon, we will treat the pillars as if they were perfect rotated half-wave plates. In this ideal case, the Jones pupil matrix can be expressed as follows: 
\begin{equation}
\textbf{J}(x,y) = 
\begin{bmatrix}
    \cos2\theta(x,y) & \sin2\theta(x,y)\\
    \sin2\theta(x,y) & -\cos2\theta(x,y)
\end{bmatrix}
\end{equation}
The elements of the amplitude response matrix are defined in accordance with (\ref{eq:ARM}):
\begin{equation}
\begin{aligned}
\text{ARM}_{XX}&=\iint \cos2\theta(x,y)\thinspace e^{-2i\pi(xf_x+yf_y)}\thinspace dxdy\\
\text{ARM}_{XY}&=\iint \sin2\theta(x,y)\thinspace e^{-2i\pi(xf_x+yf_y)}\thinspace dxdy\\
\text{ARM}_{YX}&=\text{ARM}_{XY}\quad\text{;}\quad\text{ARM}_{YY}=-\text{ARM}_{XX}
\end{aligned}
\end{equation}
Thus, the PSF will be proportional to the sum of square modulus of ARM elements, which are the results of the Fourier transform of the sine or cosine of $2\theta(x,y)$. This clearly shows that there is a strong link between PSF and the orientation distribution of the pillars in the metasurface.

\begin{figure*}
\includegraphics[width=0.9\linewidth]{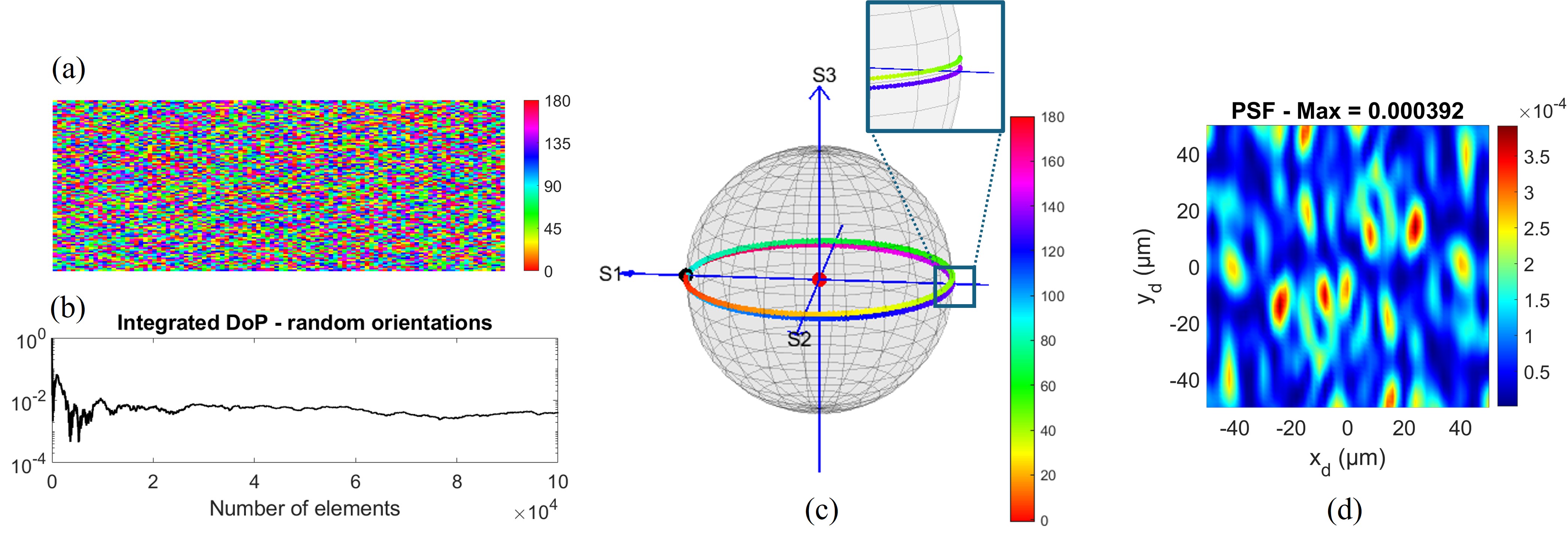}
    \caption{(a) Spatial distribution of pillar orientations (in degrees), randomly distributed across a $100\text{x}100$ elements device  - (b) $\text{DoP}_\Omega$ as a function of the number of pillars. - (c) Possible local output states at $\lambda=1605$ nm for pillar orientations $\theta$ ranging from $0^\circ$ to $180^\circ$. The red point represents their spatial integration and the blue point represents the incident polarization. - (d) Resulting PSF at $\lambda=1605$ nm.}
    \label{fig:RandomResults}
\end{figure*} 

\subsection{Implementation of the scrambling function}
\label{sec:ScramblingFunctionImplementation}

Spatial depolarization is defined by the reduction of the integrated Degree of Polarization (DoP$_\Omega$), written as \cite{huard_polarisation_1994}
\begin{equation}
    \text{DoP}_\Omega = \frac{\sqrt{\langle S_1(x,y) \rangle ^2+ \langle S_2(x,y) \rangle^2 + \langle S_3(x,y) \rangle^2}}{\langle S_0(x,y) \rangle}
\end{equation}
where the $S_i$ coefficients are the transformation of the output field $\ket{E_{out}(x,y)}$ into Stokes parameters. $\langle \rangle$ represents the spatial integration of these parameters over the spatial domain $\Omega$.
To ensure spatial depolarization, we need $\text{DoP}_\Omega$ to be as close to 0 as possible. This is achieved by building a metasurface with pillars having the optimized shape described in Section \ref{sec:UnitCellDesign} but varying orientations $\theta$, as shown in Fig.\ref{fig:Metasurface}. In the same way as a half-wave plate would, pillars will locally rotate the incident polarization state by an angle of $2\alpha$, $\alpha$ being the angle between the pillar fast axis and the incident polarization direction. For a same incident state, pillars with different orientation angle will produce different output polarization states. The distribution of pillars orientation over the metasurface $\theta(x,y)$ must then be chosen so the spatial integration of all the local output states gives a minimized DoP$_\Omega$. Depolarization performance is not the only factor that must be considered; the choice of orientation distribution $\theta(x, y)$ must also be dictated by the minimization of the PSF degradation. As outlined in Section \ref{sec:FourierMatrixFormalism}, these two criteria are not entirely independent of each other. Therefore, the challenge is to find the configuration that best satisfies both.

\section{Results}
\subsection{Random Pillar Orientation}
\label{sec:RandomPillarOrientation}

An intuitive way to scramble polarization is to randomly distribute the pillar orientations over the metasurface \cite{hu_all-dielectric_2020}. Each pillar will have a different orientation angle $\theta$, randomly chosen in the interval [$0^\circ$;$180^\circ$] (Fig.\ref{fig:RandomResults}a), and will generate a different local polarization state.

Possible local output states are represented on the Poincaré sphere in Fig.\ref{fig:RandomResults}c, they are close to the equator thanks to the half-wave plate behavior of the pillars. The barycenter of these points, which represents the spatial integration of the different local states, is very close to the center of the sphere, indicating the depolarizing properties of the device.
If we consider a metasurface of dimensions $17\thinspace\text{mm}\times 5.5\thinspace\text{mm}$ to match MicroCarb pupil size and a pillar periodicity $P$ of 0.8 µm, it represents approximately 150 million pillars, making this random distribution highly efficient in terms of depolarization performance (Fig.\ref{fig:RandomResults}b).  However, this random organization has a terrible impact on the image quality. The resulting PSF is shown in Fig.\ref{fig:RandomResults}d, where the intensity scale must be noted. The values were normalized using the central intensity of the ideal PSF, obtained without the metasurface.

When the randomly organized metasurface (Fig.\ref{fig:RandomResults}a) is considered at the entrance pupil of an imaging system, the PSF becomes a speckle pattern (Fig.\ref{fig:RandomResults}d). As our study focuses on imaging applications, a better trade-off between depolarization and image quality needs to be investigated.

\subsection{Linearly Varying Orientation}
\label{sec:LinearlyVaryingOrientation}
\begin{figure*}
    \centering
    \includegraphics[width=\linewidth]{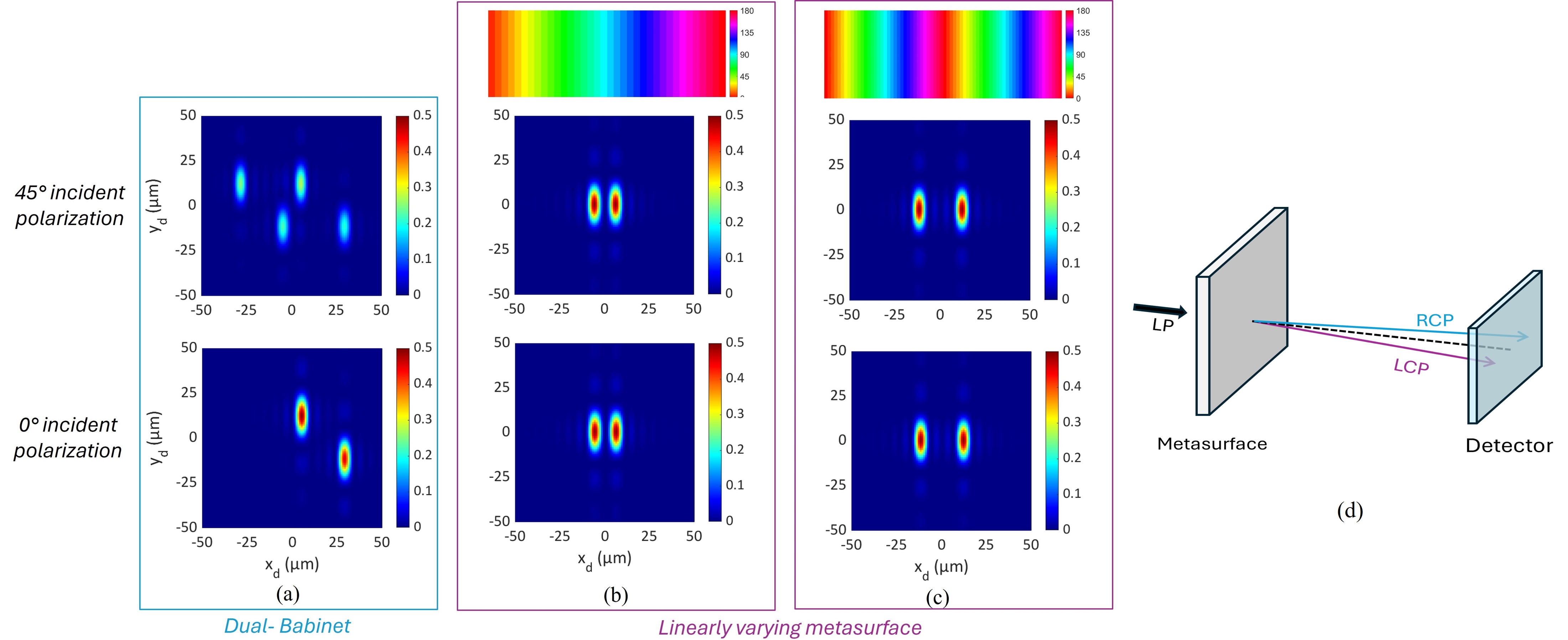}
    \caption{PSFs obtained at $\lambda=1605$ nm in the Microcarb configuration for both $45^\circ$ and $0^\circ$ linear incident polarization states, when using either a Dual-Babinet scrambler (a) or a metasurface with linearly varying orientations [(b) and (c)]. The distributions of pillar orientations (in degrees) used across the different metasurfaces are shown at the top of the corresponding PSFs: (b) $d\theta$ = $5^\circ$, $N_c = 1$, (c) $d\theta$ = $5^\circ$, $N_c = 2$. - (d) The metasurface applies opposite phase slopes to circular polarizations: incident linear polarization is split into its circular components, resulting in 2 spots on the detector.}
    \label{fig:PSF_lin}
\end{figure*}

As mentioned in the previous section, we want our component to have minimal impact on the system PSF, in addition to optimal scrambling performance. A major problem with prism-based scramblers is their detrimental effect on image quality. The PSF of systems using this type of scrambler is altered by the so-called "diamond effect" \cite{loesel_microcarb_2018,chipman_analysis_1990}, which splits the PSF into multiple spots on the detector with varying flux proportions (see Fig.\ref{fig:PSF_lin}a). Therefore, it is important to design a polarization scrambling metasurface that reduces these parasitic effects.

\begin{figure}[H]
    \centering
    \includegraphics[width=\linewidth]{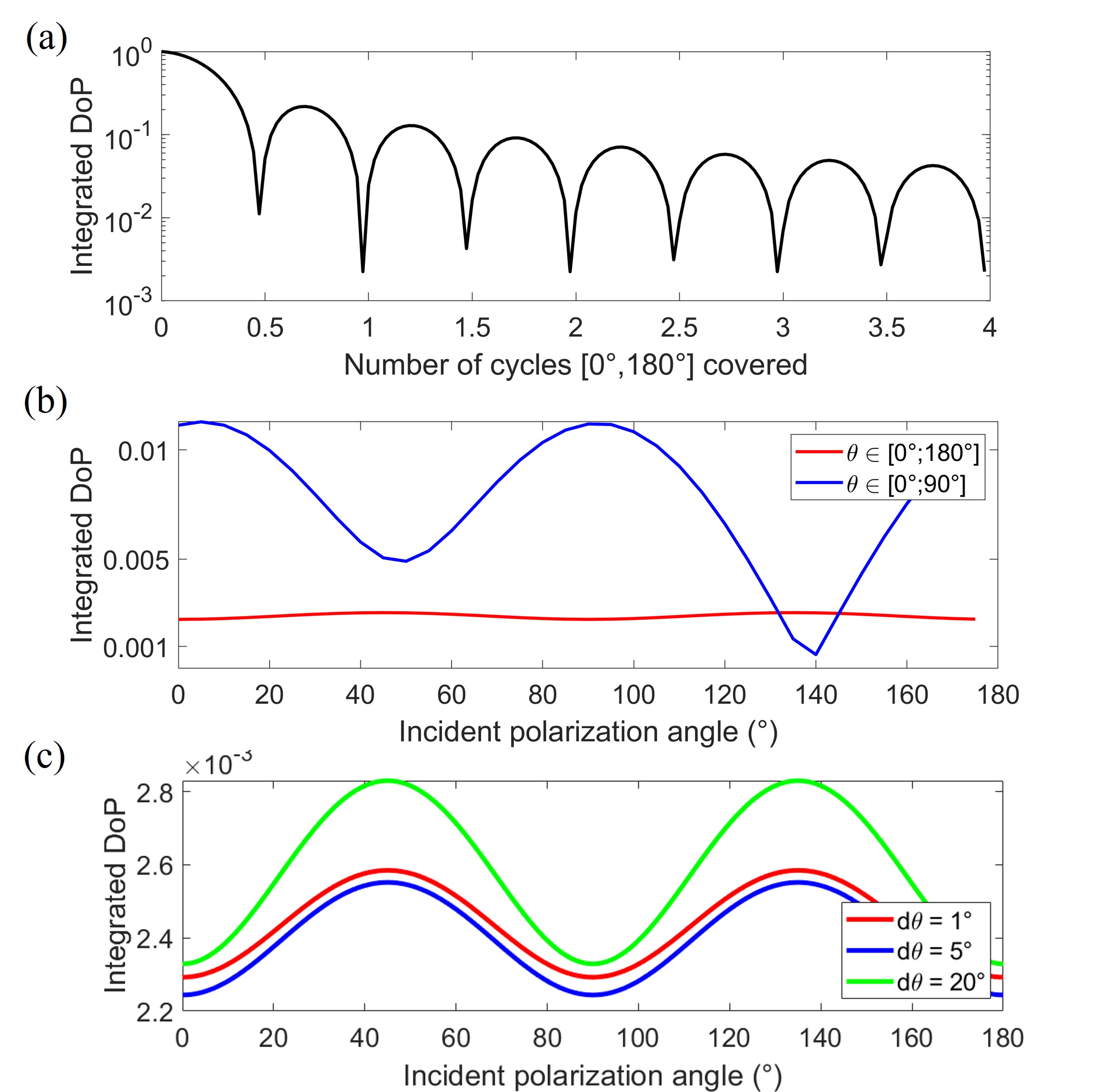}
    \caption{(a) DoP$_\Omega$ as a function of pillar orientation coverage across the metasurface, for a 0° linear incident polarization. - (b) Stability of DoP$_\Omega$ against the direction of the linear incident polarization for configurations $\theta\in[0^\circ, 180^\circ]$ (red) and $\theta\in[0^\circ,90^\circ]$ (blue). - (c) Influence of the orientation step $d\theta$ on the stability of DoP$_\Omega$ [Peak-to-valley ratios: 1.128 (1°), 1.137 (5°), and 1.215 (20°)].}
    \label{fig:Lin_Distribution}
\end{figure}

One possible solution is to use pillars with linearly varying orientations in one direction of the metasurface. The total coverage represents the range of pillar orientations across the component. Fig.\ref{fig:Lin_Distribution}a shows that DoP$_\Omega$ is minimized when this total coverage contains an integer number of cycles $[0^\circ,90^\circ]$. However, since pillars are not perfect half-wave plates, it is best to consider an integer number of cycles $[0^\circ,180^\circ]$ to ensure stable scrambling performance regardless of the direction of incident polarization (see Fig.\ref{fig:Lin_Distribution}b). Moreover, due to fabrication constraints, the variation in pillars orientations can not be continuous. Therefore, we need to consider an orientation step $d\theta$. The metasurface will be divided into vertical stripes, each containing pillars with identical orientations. The orientation step, $d\theta$, is the difference in orientation between the pillars of adjacent stripes. Because the pillars are not perfect half-wave plates for every $\theta$, the integrated DoP$_\Omega$ has a slight dependency on the direction of incident polarization. This dependency can be quantified by the peak-to-valley ratios of the curves in Fig.\ref{fig:Lin_Distribution}c. This ratio decreases as the orientation step $d\theta$ is reduced. For the remainder of this work, we will use value of $d\theta=5^\circ$, which constitutes a good balance between stability of DoP$_\Omega$ and manufacturing feasibility.

Figure \ref{fig:PSF_lin} shows the resulting PSF patterns when using either a Dual-Babinet or a metasurface with a linear distribution of pillar orientations. In both cases, the scrambler is located at the entrance pupil of the optical system. In the case of the metasurface, we observe a splitting of the PSF into two spots of equal intensity, corresponding to the circular components of the linear incident state. Indeed, since the pillars are optimized as local half-wave-plates, they apply opposite Pancharatnam-Berry phases to right- and left-handed circular polarization states. By linearly distributing their orientations, we induce opposite phase slopes for circular polarizations, resulting in opposite shifts in their contributions to the PSF, as illustrated in Fig.\ref{fig:PSF_lin}d. Since the illumination is considered to be linearly polarized, this division into two spots of equal intensity is obtained regardless of the direction of incident polarization. This is not the case for the Dual-Babinet scrambler, which splits the input beam on a linear polarization basis, resulting in an unstable PSF pattern (see Fig.\ref{fig:PSF_lin}a). Our new solution thus stabilizes the energy barycenter at the center of the detector and solves the geolocation error issue encountered by the prism-based scramblers \cite{loesel_microcarb_2018}. Additionally, this division into circular components remains the same regardless of the orientation step $d\theta$.

The distance between the two spots of the resulting PSF is defined by the phase slope induced by the metasurface. The induced phase slope is directly affected by the number of cycles $[0^\circ,180^\circ-d\theta]$ covered. For components of the same dimensions, the greater the number of cycles in the pupil width, the greater the applied phase slope, resulting in a greater distance between the two spots, as shown in Fig.\ref{fig:PSF_lin}c. The optimal configuration, in which the two spots are as close together as possible is shown in Fig.\ref{fig:PSF_lin}b and is obtained by covering only one cycle over the metasurface. 

The phase slope applied by the metasurface also depends on the device dimensions. Using a larger component reduces the generated phase slope and thus the spacing between the two spots. In our case, we want the global dimensions of the metasurface to match those of the instrument pupil. For the MicroCarb instrument, the pupil has rectangular dimensions of 17 mm x 5.5 mm, which explains the elongated shape of the resulting PSF in Fig.\ref{fig:PSF_lin}.

Although our solution produces a PSF divided into two spots, the distance $d$ between them is relatively small. It can be expressed in terms of the wavelength $\lambda$, the focal length $f$, the pupil width $L_x$, and the number of cycles $[0^\circ,180^\circ-d\theta]$ covered $N_c$ as follows
\begin{equation}
    d = \frac{2\lambda fN_c}{L_x}
\end{equation}
The PSFs shown in Fig.\ref{fig:PSF_lin} are obtained with numerical simulations using the real parameters of the MicroCarb instrument ($L_x$ = 17 mm and $f$ = 63 mm), and the spots spacing obtained with the optimal configuration shown in Fig.\ref{fig:PSF_lin}b is about $d = 11$ µm, which is smaller than the pixel size of the MicroCarb instrument (15 µm), when traditional Dual-Babinet scramblers divide the PSF up to four spots spread over several pixels of the detector (Fig.\ref{fig:PSF_lin}a) \cite{loesel_microcarb_2018}. Our solution reduces both the number of spots and their spacing while stabilizing the energy barycenter. These improvements are significant compared to the diamond effect induced by prism-based scramblers (see Fig.\ref{fig:PSF_lin}).

\begin{figure}[H]
    \centering
    \includegraphics[width=\linewidth]{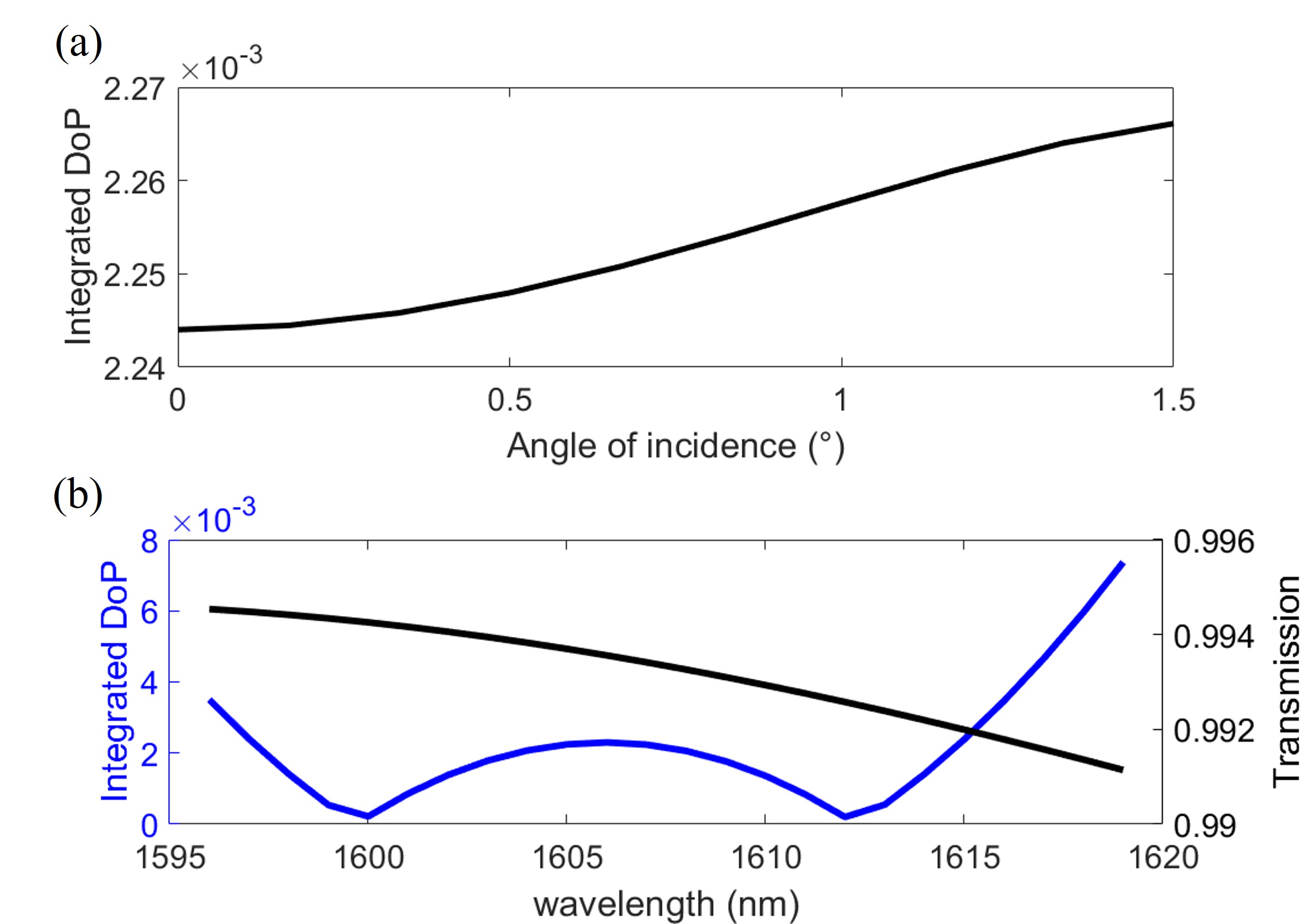}
\caption{(a) Angle of incidence dependence of DoP$_\Omega$ at $\lambda = 1605$ nm. - (b) Wavelength dependence of DoP$_\Omega$ and global transmission over the MicroCarb B2 band for a 100$\%$ $0^\circ$ linearly polarized input beam.}
\label{fig:WavelengthDependence}
\end{figure}
Figure \ref{fig:WavelengthDependence}a illustrates that depolarization performance is also maintained at oblique incidence. Our analysis is limited to the MicroCarb field of view, which is less than $\pm 1^\circ$ in the instrument pupil \cite{pasternak_microcarb_2017}.
Figure \ref{fig:WavelengthDependence}b shows the wavelength dependence of the simulated DoP$_\Omega$ obtained over one of the wavebands of interest of the MicroCarb instrument $\lambda \in [1595\text{ nm}; 1619\text{ nm}]$ \cite{pasternak_microcarb_2017}. Considering a 100$\%$ linearly polarized input beam, the global transmission exceeds 99$\%$ and the resulting output DoP$_{\Omega}$ remains below $10^{-2}$  over the entire band. This indicates spatially depolarized light and validates this first proof of concept. 

Lastly, it should be noted that the local periodic approximation, which was used to simulate the response of a single element is no longer valid when considering pillars with different orientations next to each other, as it is the case in the final metasurface design. However, it should be noted that our design uses vertical stripes with slowly linear orientation changes. This produces situations that closely resemble the periodic condition approximations used in RCWA calculations. Indeed, as most of the pillars will be surrounded by identical neighbors, the local periodic approximation is highly realistic for all the pillars located inside the stripes. Pillars located at the border of two adjacent stripes will be surrounded by pillars differing in orientation by 5° in one direction. While this makes the local periodic approximation less accurate, the change in orientation is very small and occurs in only one of the component dimensions. Therefore, we believe that the large number of pillars for which the periodic approximation is valid will offset this parasitic effect.

\section{Discussion}
\subsection{Multi-band configuration}
\begin{figure}[!htpb]
    \centering
    \includegraphics[width=\linewidth]{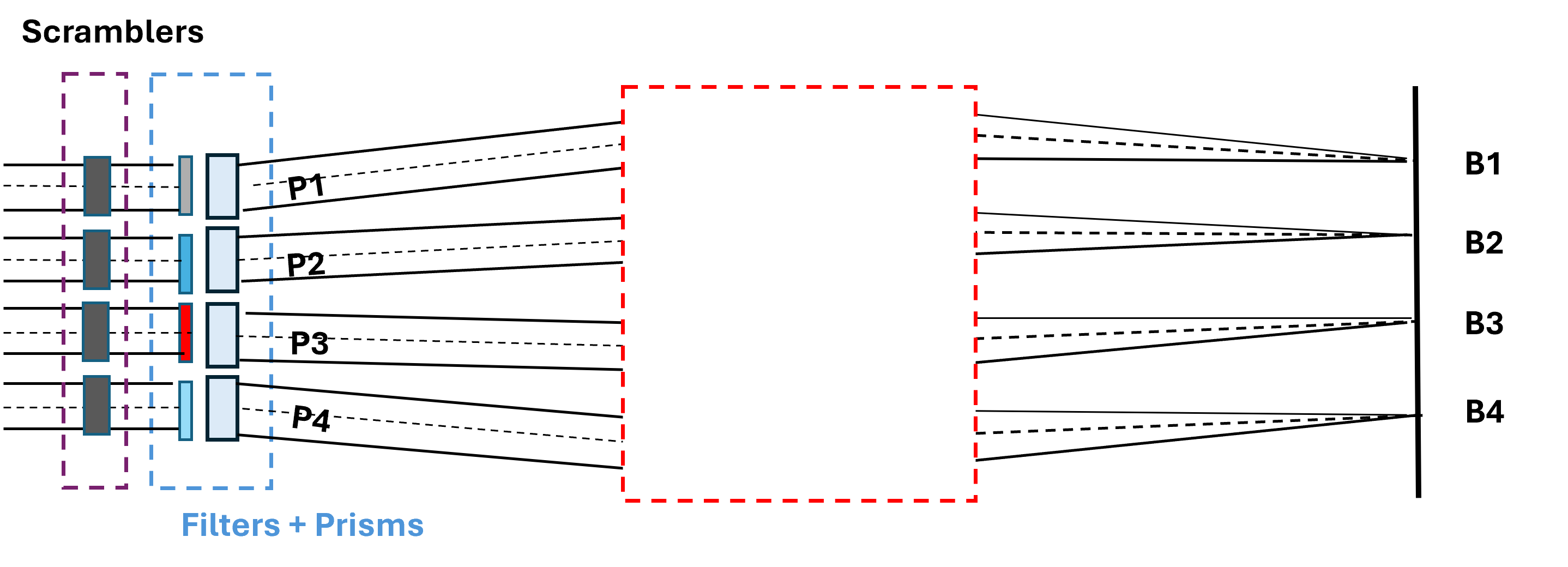}
    \caption{MicroCarb sub-pupils organization}
    \label{fig:MicroCarbDesign}
\end{figure}

The MicroCarb instrument implements four distinct spectral bands labeled B1 to B4 through four sub-pupils \cite{pasternak_microcarb_2017}, as shown in Fig. \ref{fig:MicroCarbDesign}. These four channels operate independently, with each band's wavelength range defined by a specific optical interference filter. The Dual-Babinet polarization scrambler is placed in front of these four sub-pupils, making this device much larger than the instrument's pupil dimensions.

Our solution involves installing four metasurfaces, each optimized for a specific band and placed at the corresponding sub-pupil. This approach offers significant advantages in terms of simplicity and compactness. Moreover, it solves the chromaticity issue of metasurface unit cell behavior, which leads to the progressive degradation of metasurface polarization scrambler performance when the operating wavelength is too far from the design wavelength.

The optimization process described in Section \ref{sec:UnitCellDesign} for band B2 is applied to the MicroCarb B1, B3 and B4 wavebands. Optimized geometric parameters can be defined for each channel, enabling the metasurface unit cell to behave as a half-wave plate in each case. The optimized unit cell parameters for the different wavebands are listed in Table \ref{tab:shape-functions}.
\begin{table}[htpb]
\caption{Unit cells optimized parameters for the wavebands of the MicroCarb instrument}
  \label{tab:shape-functions}
  \centering
\begin{tabular}{|c|c|c|c|c|c|c|}
\hline
 Band & $\lambda$ (nm) & Pillar & $D_x$ (nm) & $D_y$ (nm) & $P$ (nm) & $h$ (nm) \\
\hline
B1 & 758 - 769 & TiO$_2$ & 106 & 288 & 500 & 1000\\
B2 & 1596 - 1619 & Si  & 218 & 400 & 800 & 1000\\
B3 & 2023 - 2051 & Si & 269 & 529 & 1000 & 1200\\
B4 & 1264 - 1282 & Si & 141 & 313 & 650 & 1100\\
\hline
\end{tabular}
\end{table}

\subsection{Partially polarized illumination}
All of the results presented above were obtained by considering fully polarized illumination. This final section will therefore study the impact of a partially polarized incident state on the performance of the metasurface. This partial polarization can be accounted for using a Mueller matrix extension of the formalism described in Section \ref{sec:FourierMatrixFormalism}. The instrument's imaging performance can be evaluated using the point spread matrix \textbf{PSM} \cite{breckinridge_polarization_2015}, which is the transformation of the \textbf{ARM} matrix into a $4\times 4$ Mueller matrix using the following equation \cite{chipman_book_2018}:
\begin{equation}
\textbf{PSM}=\textbf{U}(\textbf{ARM}\otimes\textbf{ARM}^*)\textbf{U}^{-1}
\end{equation}
where
\begin{equation}
\textbf{U}=\frac{1}{\sqrt{2}}
\begin{bmatrix}
1 & 0 & 0 & 1\\
1 & 0 & 0 & -1\\
0 & 1 & 1 & 0\\
0 & -i & i & 0
\end{bmatrix}
\end{equation}
By combining this \textbf{PSM} with the illumination Stokes vector $\ket{S_{in}}$, we get a resulting Stokes vector $\ket{S_{out}}$ whose first coefficient $S_0$ give access to the PSF.
\begin{equation}
\ket{S_{out}(x_d,y_d)}=\begin{pmatrix} S_0 \\S_1 \\ S_2 \\ S_3\end{pmatrix}=\textbf{PSM} \ket{S_{in}}
\label{eq:PSM}
\end{equation}
where the illumination $\ket{S_{in}}$ is partially linearly polarized, with an incident $\text{DoP}_{in}$ between 0 and 1.

\begin{figure}[H]
    \centering
    \includegraphics[width=\linewidth]{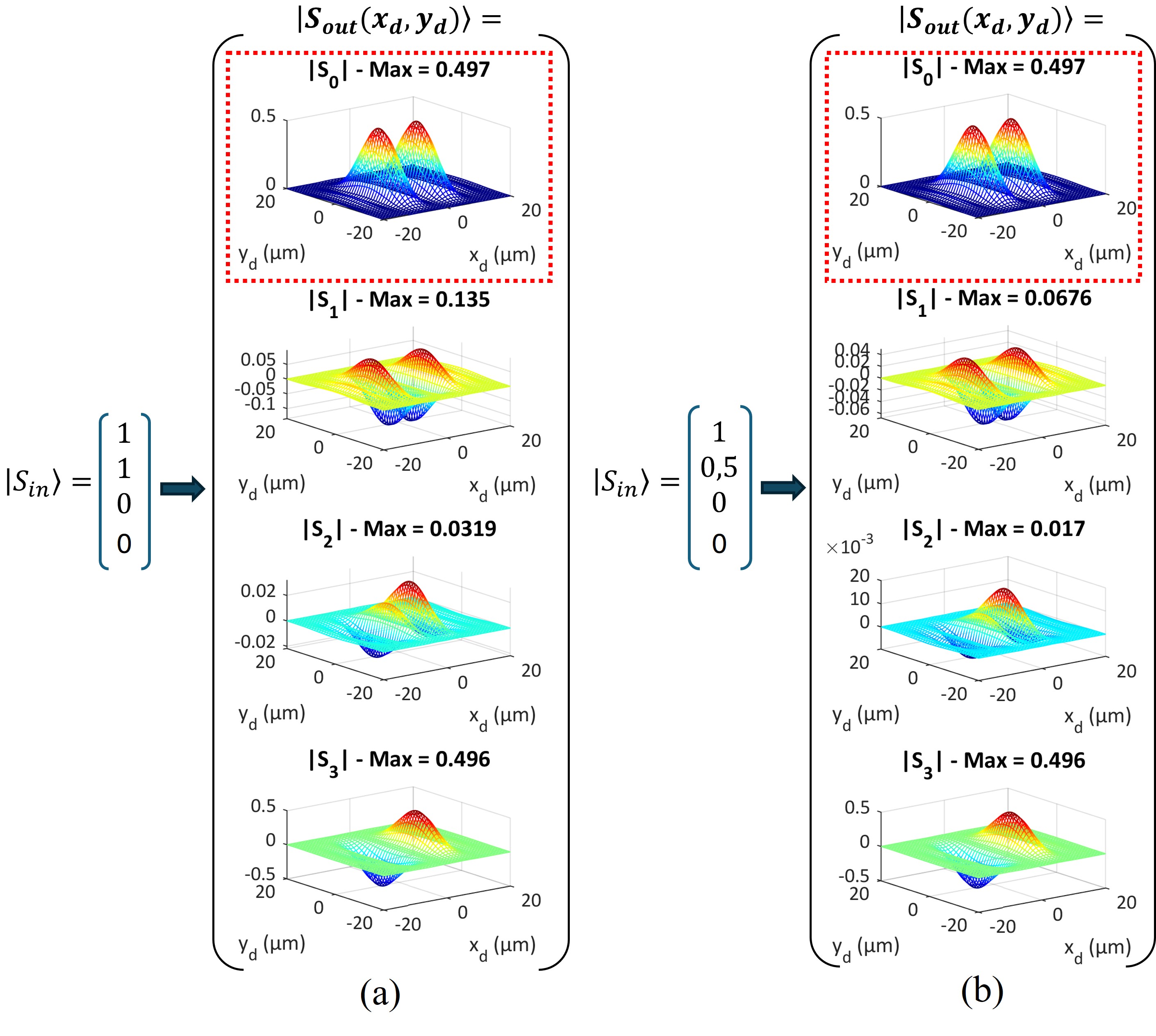}
    \caption{Stokes vectors $\ket{S_{out}(x_d,y_d)}$ obtained for linearly polarized illumination at $0^\circ$ with (a) DoP$_{in}=1$, (b) DoP$_{in}=0.5$; $S_0$ components circled in red represent the resulting PSF.}
    \label{fig:Stokes}
\end{figure}

Figure \ref{fig:Stokes} shows the resulting Stokes vectors $\ket{S_{out}(x_d,y_d)}$ obtained for a $0^\circ$ linearly polarized illumination and different values of DoP$_{in}$. In both cases, the spatial dependence of the $S_0$ coefficient, and consequently the resulting PSF are identical (circled in red in Fig.\ref{fig:Stokes}). The other three Stokes coefficients can be used to deduce the polarization decomposition of the resulting PSF.

\begin{figure}[H]
    \centering
    \includegraphics[width=0.95\linewidth]{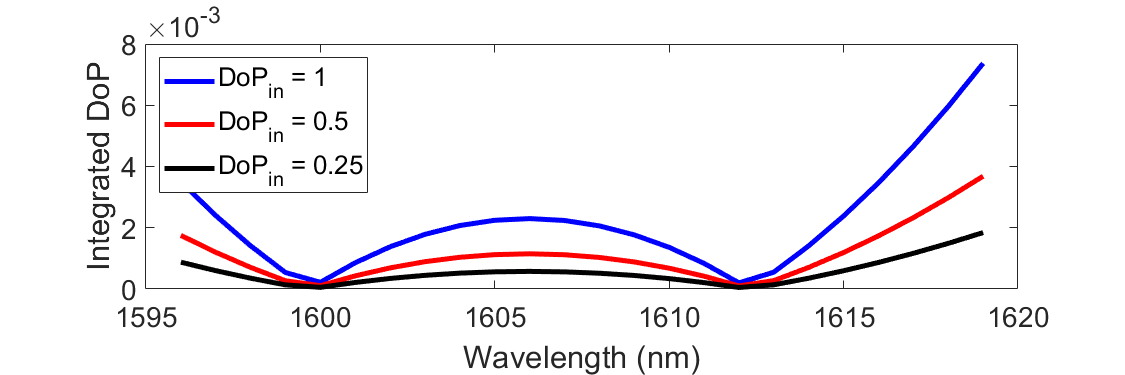}
    \caption{ DoP$_{\Omega}$ obtained over the B2 waveband, for a $0^\circ$ linearly polarized incident state and different values of DoP$_{in}$.}
    \label{fig:DoP_partial}
\end{figure}

Figure \ref{fig:DoP_partial} shows that depolarization performance is not negatively impacted by considering partially polarized illumination. The $100\%$ polarized illumination considered in all previous sections is actually the worst-case scenario for achieving maximal depolarization.

\section{Conclusion}
We designed a novel polarization scrambling solution that uses a metasurface. We implemented anisotropic dielectric scatterers with optimized dimensions that act as half-wave plates and linearly varying orientation across the width of our device. Our device minimizes the integrated DoP and reduces the impact on the system image quality while offering a greater compactness than prism-based scrambler. This new solution could offer a simpler and more efficient alternative to existing scramblers. In particular, prism-based scramblers require a high level of design and manufacturing complexity to mitigate variations in prism birefringence caused by stress changes applied by the component's mechanical holder under thermal loading \cite{lopez_manufacturing_2021}. Replacing the prism assembly with a single metasurface, whose birefringence is solely determined by its pillar parameters, could reduce the complexity of the component integration.

We are currently manufacturing this design to produce a full-size prototype that matches the Microcarb pupil dimensions (i.e., 17 mm x 5.5 mm). This will demonstrate the feasibility of this scrambling solution for realistic sample sizes used in space applications.

\begin{acknowledgments}
This work has been achieved thanks to the support of the French Centre National d’Etudes Spatiales - CNES, and the company CILAS in the context of the joint laboratory LabTOP
\end{acknowledgments}

\appendix*
\section{Local contribution of a rotated pillar}
\label{sec:LocalContributionofaPillar}
The results presented in Section \ref{sec:UnitCellDesign} have been obtained for a pillar's orientation $\theta$ of $0^\circ$. As a first approximation, the Jones matrix $\textbf{J}(\theta)$ of a rectangular pillar rotated by an angle $\theta$ can be defined by the following equation, as explained in \cite{arbabi_dielectric_2015}:
\begin{equation}
\textbf{J}(\theta) = \textbf{R}(-\theta)\thinspace\textbf{J}\thinspace\textbf{R}(\theta)
\label{eq:J_Theta}
\end{equation}
involving the rotation matrices $\textbf{R}(\theta)$ and $\textbf{R}(-\theta)$.

\begin{figure*}
     \centering
     \includegraphics[width=0.9\linewidth]{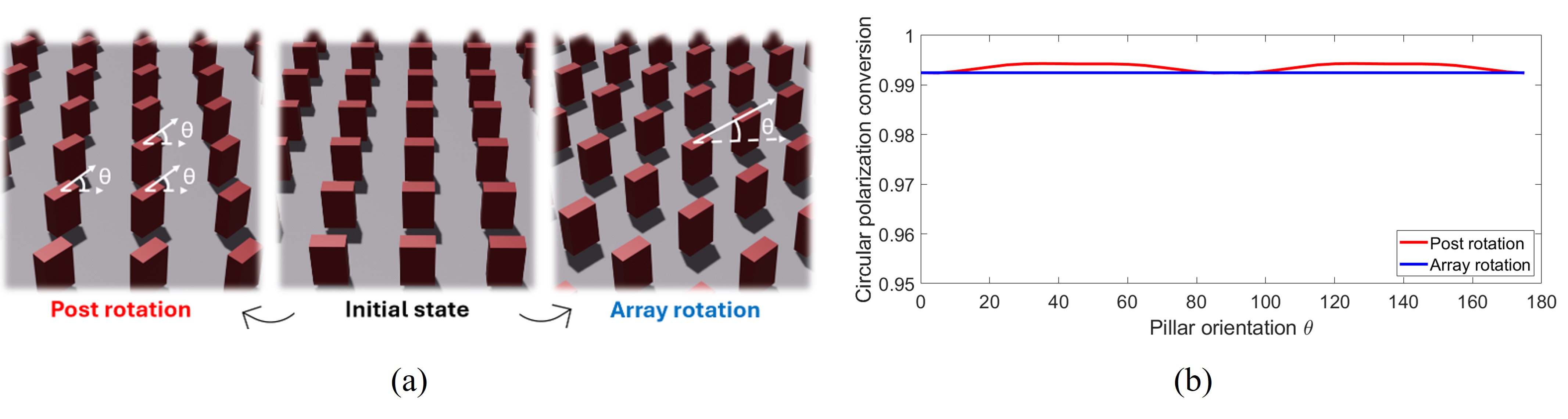}
     \caption{(a) Visualization of the two different approaches that can be used to model the rotation of pillars: "Post rotation" [RCWA simulations were carried out for each angle $\theta$] and "Array rotation" [only one RCWA simulation was carried out at $\theta = 0^\circ$, and the behavior at other angles was found using Eq.\ref{eq:J_Theta}] - (b) Circular polarization conversion efficiency $\eta$, evaluated using the two approaches, as a function of pillar orientation $\theta$ for a $0^\circ$ linearly polarized incident state.}
     \label{fig:Rotation}
\end{figure*}

This expression for $\textbf{J}(\theta)$ models the rotation of an entire array of identical pillars by an angle $\theta$, which is not exactly the same as rotating each pillar by an angle $\theta$ around its own axis, the difference being illustrated in Fig.\ref{fig:Rotation}.

In the first case, represented by equation (\ref{eq:J_Theta}), the entire array is rotated and the arrangement of the scatterers within the metasurface as well as the interactions between adjacent meta-atoms remain the same. This type of pillar rotation does not affect the circular polarization conversion efficiency $\eta$. Conversely, when considering the rotation of all posts around their own axes, the spacing between adjacent pillars is impacted, as shown in Fig.\ref{fig:Rotation}, resulting in a slight variation in the circular polarization conversion efficiency (Fig.\ref{fig:Rotation}). These variations in behavior are relatively small, so Eq.(\ref{eq:J_Theta}) is suitable in most cases. However, our application requires a highly precise calculation of the DoP, which is a very sensitive parameter that considers the pillars collectively. Therefore, we choose to consider these small variations in the half-wave plate behavior of rotated pillars. To accomplish this, we will not use Eq.(\ref{eq:J_Theta}). Instead, we will run an RCWA simulation for each angle $\theta$ considered.

\end{document}